\documentclass[aps,preprint,showpacs]{revtex4-1}

\usepackage{bm}
\usepackage{graphicx}
\usepackage{latexsym}
\usepackage{amsmath}
\usepackage[colorlinks=true,citecolor=blue,linkcolor=blue]{hyperref}
\usepackage{bbding}
\usepackage[caption=false]{subfig} 
\usepackage{float}
\usepackage{subfloat}
\usepackage[section]{placeins}
\usepackage{csquotes}
\usepackage{relsize}
\usepackage{enumitem}
\DeclareGraphicsExtensions{.pdf,.png,.jpg}

\LetLtxMacro{\OldSqrt}{\sqrt}
\newcommand{\ClosedSqrt}[1][\hphantom{3}]{\def\DHLindex{#1}\mathpalette\DHLhksqrt}
\makeatletter
    \newcommand*\bold@name{bold}
    \def\DHLhksqrt#1#2{%
        \setbox0=\hbox{$#1\OldSqrt{#2\,}$}\dimen0=\ht0\relax%
        \advance\dimen0-0.2\ht0\relax
        \setbox2=\hbox{\vrule height\ht0 depth -\dimen0}%
        {\hbox{$#1\expandafter\OldSqrt\expandafter[\DHLindex]{#2\,}$}
        \lower\ifx\math@version\bold@name0.6pt\else0.4pt\fi\box2}
    }
    \renewcommand*{\sqrt}[2][\ ]{\ClosedSqrt[\leftroot{-2}\uproot{1}#1]{#2}\kern0.1em} 
\makeatother

\renewcommand\vec{\mathbf}

\begin{document}

\title{Quasi-monoenergetic Laser Plasma Positron Accelerator \\ using Particle-Shower Plasma-Wave Interactions}

\author{Aakash A. Sahai}
\affiliation{Department of Physics \& John Adams Institute for Accelerator Science, Blackett Laboratory, Imperial College London, SW7 2AZ, United Kingdom\vspace{7.5mm}}
\email[corresponding author: ~]{aakash.sahai@gmail.com}

\begin{abstract}
An all-optical centimeter-scale laser-plasma positron accelerator is modeled to produce quasi-monoenergetic beams with tunable ultra-relativistic energies. A new principle elucidated here describes the trapping of divergent positrons that are part of a laser-driven electromagnetic shower with a large energy spread and their acceleration into a quasi-monoenergetic positron beam in a laser-driven plasma wave. Proof of this principle using analysis and Particle-In-Cell simulations demonstrates that, under limits defined here, existing lasers can accelerate hundreds of MeV pC quasi-monoenergetic positron bunches. By providing an affordable alternative to kilometer-scale radio-frequency accelerators, this compact positron accelerator opens up new avenues of research. 
\end{abstract}
\maketitle

Mono-energetic positron accelerators intrinsic to positron-electron ($e^+-e^-$) colliders at energy frontiers \cite{spear-1966-pep-1977,slc-lep-1984} have been fundamental to many important discoveries \cite{j-psi-spear, tau-lepton,BaBar-CP-violations,lep-WZ-meas} that underpin the Standard Model. Apart from high-energy physics (HEP), mono-energetic $e^+$-beams of mostly sub-MeV energies are also used in many areas of material science \cite{positron-materials,positron-book}, medicine \cite{positron-medicine} and applied antimatter physics \cite{applied-physics}. Applications have however not had ready access to positron accelerators and have had to rely on alternative sources such as $\beta^+$-decay \cite{beta-plus-decay}, (p,n) reaction \cite{pn-reactions} and pair-production \cite{EM-cascade-shower} of MeV-scale photons from - fission reactors \cite{fission-positron}, neutron-capture reactions \cite{neutron-capture-positron} or MeV-scale $e^-$-beams impinging on a high-Z target \cite{high-energy-positron}.

Positron accelerators have evidently been scarce due to complexities involved in the production and isolation of elusive particles like positrons \cite{slc-lep-1984,high-energy-positron} in addition to the costs associated with the large size of radio-frequency (RF) accelerators \cite{RF-accelerators}. The size of conventional RF accelerators is dictated by the distance over which charged particles gain energy under the action of breakdown limited \cite{breakdown-limit} tens of $\rm MVm^{-1}$ RF fields sustained using metallic structures that reconfigure transverse electromagnetic waves into modes with axial fields. This limit also complicates efficient positron production \cite{slc-lep-1984,EM-cascade-shower}, which has required a multi-GeV $e^-$-beam from a kilometer-scale RF accelerator \cite{RF-accelerators} to interact with a target. Furthermore, the positrons thus produced have to be captured in a flux concentrator, turned around and  transported back \cite{flux-concentrator-return-line} for re-injection into the same RF accelerator. 

Advancements in RF technologies have demonstrated $\rm100 ~ MVm^{-1}$-scale fields \cite{ilc-clic-design} but explorations beyond the Standard Model at TeV-scale $e^+$-$e^-$ center-of-mass energies still remain unviable. Moreover, the progress of non-HEP applications of $e^+$-beams has  been largely stagnant.

\begin{figure}[!htb]
	\begin{center}
   	\includegraphics[width=0.9\columnwidth]{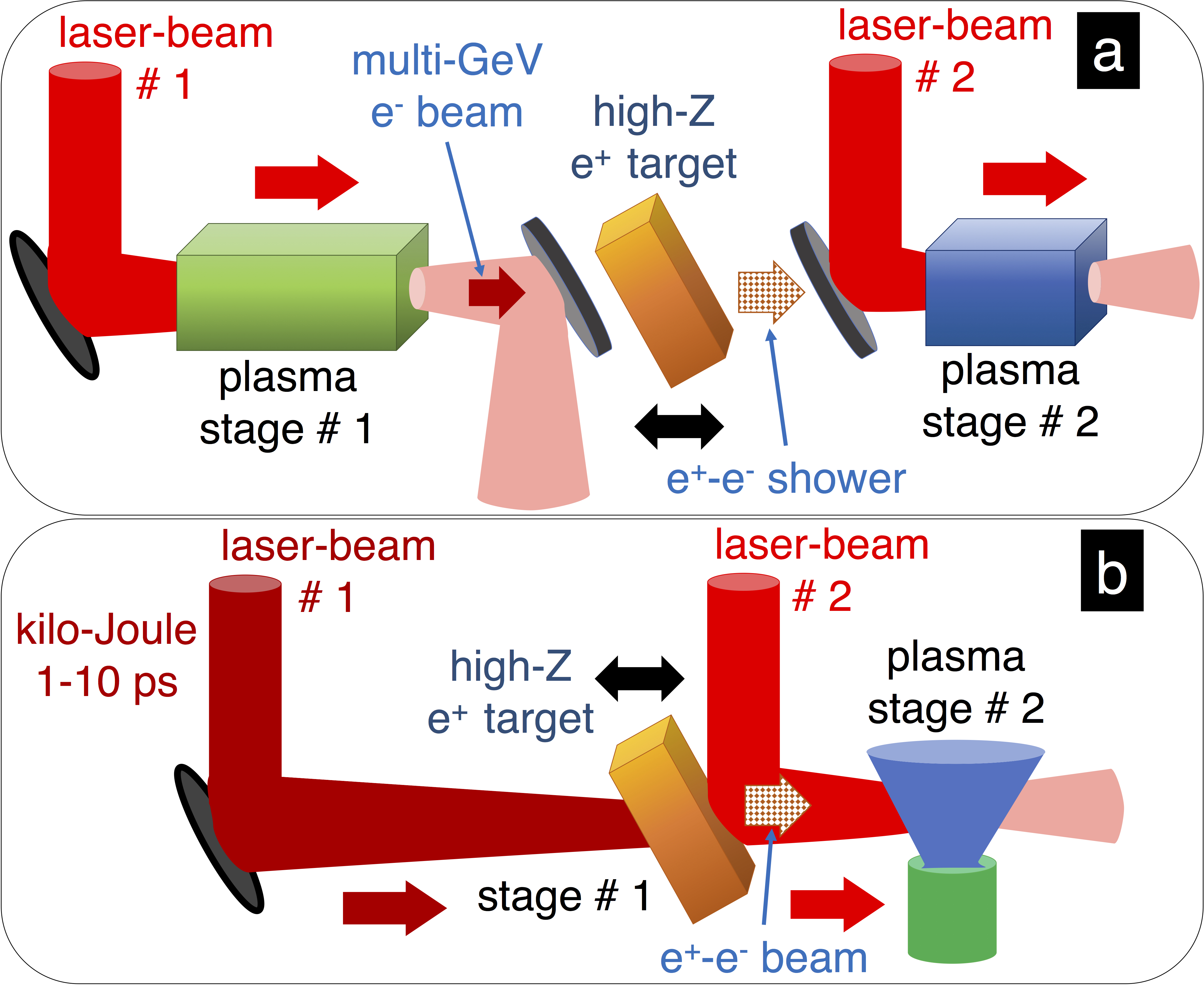}
	\end{center}
\caption{Schematic of all-optical centimeter-scale schemes of quasi-monoenergetic laser-plasma positron accelerator using the interaction of $e^+-e^-$ showers with plasma-waves.}
\label{fig:LPA-positron-schematic}
\end{figure}
\begin{figure*}[th]
	\begin{center}
   	\includegraphics[width=\textwidth]{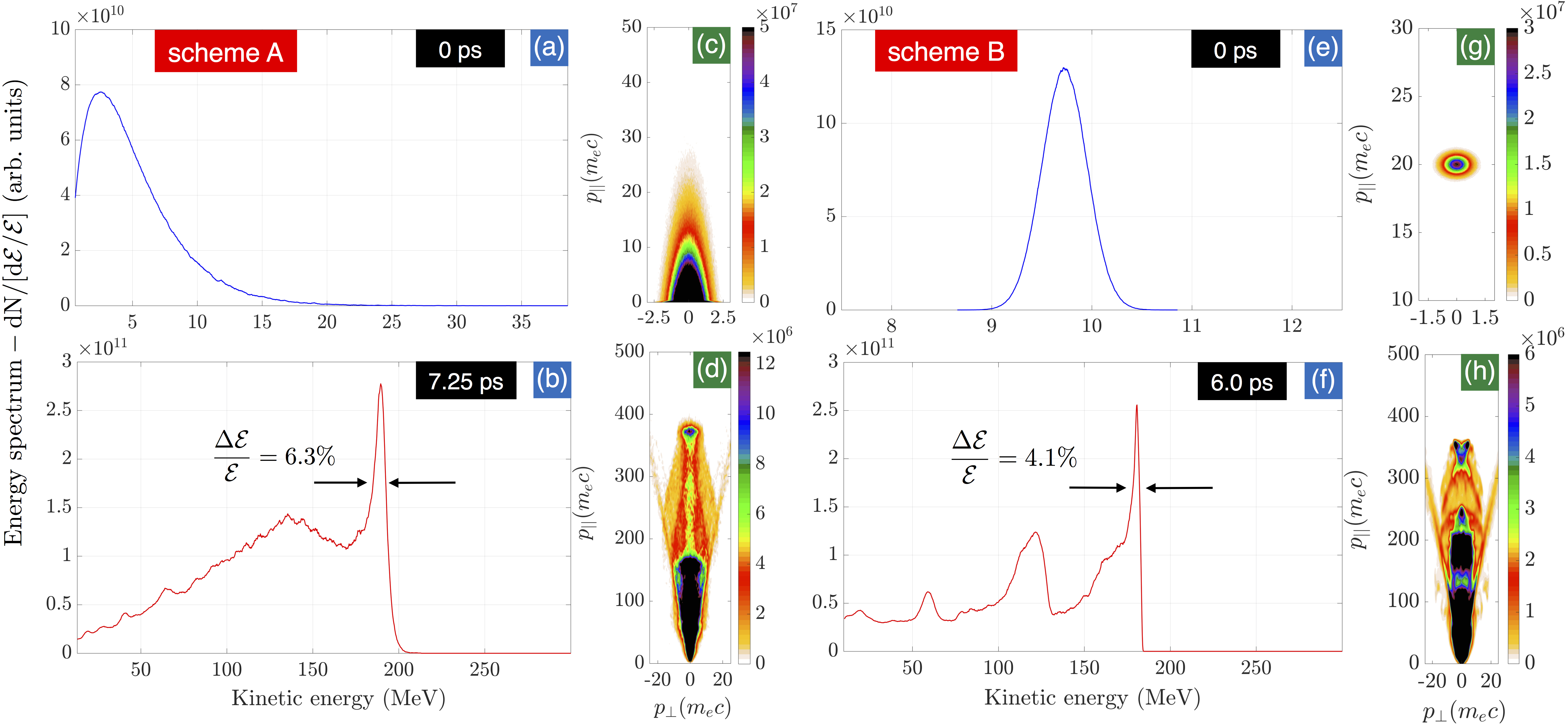}
	\end{center}
\caption{Energy spectra and $p_{\perp}-p_{\parallel}$ phase-spaces of  $e^+$-LPA accelerated $e^+$-beams modeled with $n_0=10^{18}\rm{cm^{-3}}$ using a 50fs laser with $a_0=1.4$ and Full Width at Half Maximum (FWHM) spot-size of $40\mu m$. For Scheme A (Scheme B), the initial conditions are in (a),(c) ((e),(g)) and the $e^+$-beam at 2.2 mm (1.8mm) in (b),(d) ((f),(h)).}
\label{fig:posi-energy-spectrum}
\end{figure*}
Recent efforts on compact and affordable positron accelerator design based on advanced acceleration techniques \cite{Tajima-Dawson,RMP-2009} have unfortunately been unsatisfactory. Production of $e^+-e^-$ showers using high-energy electrons from compact laser-plasma accelerator ($e^-$-LPA) \cite{Tajima-Dawson,RMP-2009,LPA-electron-accln} has been reported \cite{Sarri-shower-is-positron-beam}. However, unlike $e^+$-\enquote{beams}, showers suffer from innately exponential energy spectra. Moreover, the positron number in showers which peaks around a few MeV \cite{slc-lep-1984,leptonic-beams-Sarri}, undergoes orders-of-magnitude drop at higher energies. Another work which uses sheath fields driven by kilo-Joule (kJ) lasers in metal targets has obtained quasi-monoenergetic 10 MeV positrons \cite{posi-kj-laser} although with inherently high temperatures. Both scaling to higher energies and cooling of positrons using this mechanism is yet unexplored. Beam-driven plasma acceleration of positrons \cite{posi-slc,posi-slc-proposal} although compact by itself, depends on kilometer-scale GeV RF accelerators. Additionally, obtaining an appropriately spaced drive-witness bunch pair for beam-plasma acceleration methods is technologically difficult. 

In this letter, all-optical quasi-monoenergetic $e^+$-beam production is proposed using a centimeter-scale positron accelerator (as shown in Fig.\ref{fig:LPA-positron-schematic}). This laser-plasma positron accelerator ($e^+$-LPA) uses the interaction between laser-driven $e^+-e^-$ particle showers \cite{leptonic-beams-Sarri} and laser-driven plasma waves that support 100 $\rm \sqrt{n_0(10^{18}\rm cm^{-3})} GVm^{-1}$ fields \cite{Tajima-Dawson,RMP-2009} ($n_0$ is the plasma electron density in $\rm cm^{-3}$). This letter models the trapping of divergent positrons that are part of laser-driven particle showers and their acceleration into a quasi-monoenergetic $e^+$-beam in a laser-driven plasma wave. 

This novel compact $e^+$-LPA opens up an affordable pathway for the application of ultra-relativistic quasi-monoenergetic $e^+$-beams outside HEP as much as it invigorates research in advanced collider concepts \cite{plasma-colliders}.

The mechanism modeled in this letter uses two coupled laser-plasma interaction stages. In the first (positron-production) stage, bremsstrahlung emission from laser-driven electrons undergoes pair-production in the nuclear Coulomb field inside a high-Z target and results in an electromagnetic-cascade particle-shower \cite{EM-cascade-shower}. In scheme-A shown in Fig.\ref{fig:LPA-positron-schematic}(a), an $e^-$-LPA produces multi-GeV electrons \cite{self-guided-multi-GeV}. In scheme-B shown in Fig.\ref{fig:LPA-positron-schematic}(b), a kJ laser \cite{posi-kj-laser} produces an MeV electron flux in the pre-plasma of a solid target. The $e^+$-$e^-$ shower from the target propagates into the second (positron-acceleration) stage where a significant number of shower particles are trapped in a laser-driven plasma-wave. The fields of the plasma-wave accelerate a quasi-monoenergetic $e^+$-\enquote{beam} with typical energy spectra from particle-in-cell (PIC) simulations shown in Fig.\ref{fig:posi-energy-spectrum}(b),(f). This groundbreaking quasi-monoenergetic $e^+$-beam acceleration model defines the key principles as well as the limits of $e^+$-LPA. Recent efforts have shown that it is possible to overcome single-stage limits of electron acceleration using multistage $e^-$-LPAs albeit with a few technological challenges \cite{electron-staging-2016}.

A proof of the principle of the above described $e^+$-LPA is developed below using analysis and PIC simulations.

The first stage laser-driven $e^+-e^-$ shower are below modeled with characteristics that depend upon peak electron energy and net charge in scheme-A \cite{leptonic-beams-Sarri}, laser energy in scheme-B \cite{posi-kj-laser} in addition to the target properties. 

In scheme-A the particle-shower is modeled with an anisotropic relativistic Maxwellian distribution \cite{rel-Maxwellian-Weibel,anisotropic-rel-Maxwellian} consistent with experiments \cite{slc-lep-1984,leptonic-beams-Sarri}. This distribution in momentum space (normalized to $m_ec$), ${\bf p}=(p_{\perp},p_{\parallel})$ is
\begin{equation}
f({\bf p}) = C ~ \left(p_{\perp}^2 + p_{\parallel}^2\right) ~ {\rm exp}\left[ -\beta_{\perp} \sqrt{1 + p_{\perp}^2 + A ~ p_{\parallel}^2} ~ \right]
\label{eq:anisotropic-relativistic-maxwellian} 
\end{equation}
\noindent where $p_{\parallel}$ is along the axis of laser propagation and $p_{\perp}$ in the transverse directions, $\beta_{\perp}=m_ec^2~T_{\perp}^{-1}$, $A=T_{\parallel}T_{\perp}^{-1}$, transverse $T_{\perp}$ and longitudinal $T_{\parallel}$ temperatures are in eV and $C$  normalizes the distribution \cite{anisotropic-rel-Maxwellian}. Using experimental evidence \cite{slc-lep-1984,leptonic-beams-Sarri}, the peak particle number is at ${\rm2.3 MeV}$ ($df({\bf p})/dp_{\parallel} =  0$) with $T_{\perp} = {\rm 0.2 MeV}$ and $A = 25$. The shower positron densities here lie between $\rm 10^{15}-10^{17} cm^{-3}$ with $e^+$-to-$e^-$ density ratio ($f_{e^+} / f_{e^-}$) of between 0.1 to 0.4 \cite{leptonic-beams-Sarri}. 

Experiments on laser-driven $e^+-e^-$ showers, which observed $10^9$ positrons over 1MeV \cite{leptonic-beams-Sarri} using 0.6GeV peak energy, 100pC $e^-$-LPA electrons (with a 10J, 50fs, $\rm\lambda_0=0.8\mu m$ wavelength laser) incident on 5-10 millimeter {\it Pb} target, showed excellent agreement with Monte-Carlo particle simulations (GEANT4/FLUKA). These simulations predict many times higher $e^+$-yield \cite{Sarri-Cole-private} using multi-GeV $e^-$-LPA electrons \cite{self-guided-multi-GeV}  but the innate distribution of showers in eq.\ref{eq:anisotropic-relativistic-maxwellian} is retained.

In scheme-B, sheath-accelerated $e^+-e^-$ shower is here modeled on experiments in \cite{posi-kj-laser} that observed $10^{10}$ positrons using a 305J, $\rm\lambda_0=1.054\mu m$, $\rm\tau_p\sim10ps$ laser incident on millimeter-scale {\it Au} targets. Here this quasi-monoenergetic shower is modeled using a drifting Maxwellian distribution with a drift kinetic energy of $\rm 10 MeV$, $T_{\perp}=T_{\parallel}={\rm 200 keV}$ (isotropically) and $\rm 10^{15}-10^{16} cm^{-3}$ densities with above $f_{e^+} / f_{e^-}$ ratio.

Using the above shower models, trapping and acceleration of the shower positrons in a laser-driven plasma wave is analyzed below. The dependence of $e^+$-beam properties (energy spectrum, emittance, charge) on $e^+$-LPA second-stage parameters is also investigated.

In the electron compression phase of the wave, electron-ion charge-separation potential driven by the laser ponderomotive force \cite{Tajima-Dawson,RMP-2009} ($\propto \nabla(\rm I_0\lambda_0^2)$, where $\rm I_0$ is the peak intensity of a laser) is found to trap, focus and accelerate the shower positrons. The difference in velocity of the shower positrons and the electron compression phase ($\rm\beta_{\phi}=\left[ 1-\omega_{pe}^2/\omega_0^2 \right]^{\frac{1}{2}}$, $\rm\omega_0=2\pi c\lambda_0^{-1}$, ${\rm\omega_{pe}}=\left[4\pi n_0 \rm e^2 m_e^{-1} \right]^{\frac{1}{2}}$ is the electron plasma frequency) necessitates a careful analysis of their interaction. An analysis followed by PIC simulations below elucidates the requirements to trap shower positrons and tune the accelerated $e^+$-beam energy spectra and energy gain.

The threshold potential required to trap and retain a significant positron number can be analytically derived. The minimum kinetic energy, $\rm \mathcal{E}_{sh}=\left(\gamma_{sh}-1\right) m_ec^2$ (lab-frame momentum, $p_{\parallel}=\rm \gamma_{sh}\beta_{sh}^{\parallel}m_ec$, $\rm\gamma_{sh}=[1-\beta_{sh}^2]^{-\frac{1}{2}}$) of the positrons that are trapped is chosen to be less than the peak of the distribution in eq.\ref{eq:anisotropic-relativistic-maxwellian}. The Lorentz transformed lower-limit of trapped positron kinetic energy in wave-frame with $\gamma_{\phi}=[1-\beta_{\phi}^2]^{-\frac{1}{2}}=\omega_0/\omega_{pe}$ is
\begin{equation}
\rm \mathcal{E}'_{sh} = \left(\frac{\omega_0}{\omega_{pe}} ~ \gamma_{sh} ~ (1 - \beta_{sh}^{\parallel}\beta_{\phi}) -1 \right) ~ m_ec^2.
\label{eq:shower-energy-wave-frame} 
\end{equation}
\noindent Positrons with negative relative velocities in the wave-frame at $\rm\mathcal{E}'_{sh} $ are trapped only when a lower-limit of wave-frame potential $\Psi'$ is exceeded 
\begin{equation}
\begin{aligned}
& \rm e\Psi' \geq \mathcal{E}'_{sh}
\label{eq:trapping-condition} 
\end{aligned}
\end{equation}
\noindent Lorentz transformation of the four potential $(\Psi', \vec{A}')$ ($\vec{A}'$ is the wave vector potential) back to the lab-frame under gauge invariance gives the threshold potential $\Psi$ and $\rm  \psi_{\rm th}$
\begin{equation}
\begin{aligned}
& \rm \Psi = \frac{\omega_{pe}}{\omega_{0}} ~ \Psi' + c~\bf{A}\cdot\bm{\beta_{\phi}} \\
& \rm \psi_{\rm th} \geq  \gamma_{sh} ~ (1 - \beta_{sh}^{\parallel}\beta_{\phi}) - \frac{\omega_{pe}}{\omega_{0}}, ~ \psi = \frac{e\Psi}{m_ec^2}, ~ A_{\parallel}=0.
\label{eq:wave-potential-lab-frame} 
\end{aligned}
\end{equation}

The longitudinal trapping condition in eq.\ref{eq:wave-potential-lab-frame} is necessary but not sufficient, because particles may still transversely escape. A threshold potential is therefore necessary to constrain the divergent positrons within a transverse escape momentum contour. This potential is derived by Lorentz transforming to the shower frame at $c\beta^{\parallel}_{\rm sh}$ where the longitudinal momentum contracts and the average particle energy in the shower frame is $k_{\rm B}\rm T_{\perp}$. Thus, the threshold $\Psi''$ and $\rm\psi_{th}$ are
\begin{equation}
\begin{aligned}
& e\Psi'' \geq \alpha ~ k_{\rm B} {\rm T_{\perp}} \\
& \psi_{\rm th} \geq \alpha ~ \frac{ k_{\rm B}{\rm T_{\perp}(m_ec^2)^{-1}}}{1+\mathcal{E}_{\rm sh}{\rm (m_ec^2)^{-1}}}
\label{eq:transverse-escape-momentum}
\end{aligned}
\end{equation}
\noindent where, $\alpha > 1$ accounts for the trapping of particles away from $c\beta^{\parallel}_{\rm sh}$ in the shower momentum distribution.

The peak-shaped potential in the electron compression phase that satisfies eq.\ref{eq:wave-potential-lab-frame},\ref{eq:transverse-escape-momentum} can be modeled as $\rm \psi(\zeta,r)=$ $\rm-\psi_0 ~ {\rm sech}^2\left(\frac{\zeta-\zeta_{peak}}{L_{\Delta}}\right){\rm sech}^2\left(\frac{r-r_{peak}}{R_{\Delta}}\right) \mathcal{H}(-\psi)$, (using eq.20 in \cite{RMP-2009} and PIC data) where $\psi_0$ is the peak negative potential, $\rm\mathcal{H}$ the step function and $\zeta = z - c\beta_{\phi}t$. This potential peaks at $\rm\zeta_{peak}$ and $\rm r_{peak}$ and falls off over scale-lengths, $\rm L_{\Delta}(\psi_0)$ longitudinally and $\rm R_{\Delta}(\psi_0)$ radially. The fields in this region are both accelerating and focussing. 

The shape of the beam energy spectrum is optimized by restraining the potential (upper-limit), although eq.\ref{eq:wave-potential-lab-frame} suggests its arbitrary increase to $\psi \gg 1$, can extend trapping to $\rm p^{\parallel}_{sh}\rightarrow 0$. Upon satisfaction of the trapping condition in eq.\ref{eq:wave-potential-lab-frame},\ref{eq:transverse-escape-momentum}, this work shows that it is the profile of the potential ($\psi_0,{\rm L_{\Delta}}(\psi_0), {\rm R_{\Delta}}(\psi_0)$) which shapes the spectrum. This potential profile is dictated by the wave amplitude ($\delta n_e/n_0 = n_e/n_0 -1$ where $n_e(\zeta,r)$ is the density in the wave) in accordance with $\nabla^2 \psi = k_{\rm pe}^2 ~ \delta n_e/n_0$ \cite{RMP-2009} ($k_{\rm pe}=c^{-1}\omega_{\rm pe}/\beta_{\phi}$). Dynamics studied here shows that as the wave steepens with increasing amplitude its positron acceleration phase shrinks, ${\rm L_{\Delta}} \propto\psi_0^{-1}$. The resultant faster longitudinal field variation degrades the energy spectrum. As the positron trapping region size reduces, beam charge also decreases. A quasi-nonlinear wave with $\psi\sim\mathcal{O}(1)$ therefore turns out to be optimal.

Beam energy gain, $\rm\Delta W$ is optimized as the distance of overlap between the trapped beam and the favorable potential maximizes. This acceleration length, ${\rm L_{acc}}$ is shown to depend on the wave amplitude, $\psi_0$ and the plasma density, $n_0$. The wave amplitude is itself dictated by $n_0$ and the normalized laser vector potential, $a_0$ ($= \rm max\left(e\vec{A}_{0}/m_ec^2\right)$) in accordance with $\delta n_e/n_0 \propto k_{\rm pe}^{-2}\nabla a_0^2$, while in plasma the $a_0$ is modified by the wave density as per $\left(\nabla^2 - c^{-2}\partial^2/\partial t^2\right) a_0 = k^2_{\rm pe}(n_e/n_0) a_0$ \cite{RMP-2009}. In this work, it is found that an initially high $a_0$ or a rise in $a_0$ due to laser evolution increases the wave amplitude which shortens the potential profile and constrains ${\rm L_{acc}}$. This limit of the overlap dictates the energy gain $\rm{\Delta W = e\langle E_{\parallel} \rangle_{\rm L_{acc}} L_{acc}}$ where $\rm\langle E_{\parallel} \rangle_{\rm L_{acc}}$ ($=-\partial \psi/\partial \zeta$) is the wave longitudinal field averaged over ${\rm L_{acc}}$.

\begin{figure}[ht]
	\begin{center}
   	\includegraphics[width=0.85\columnwidth]{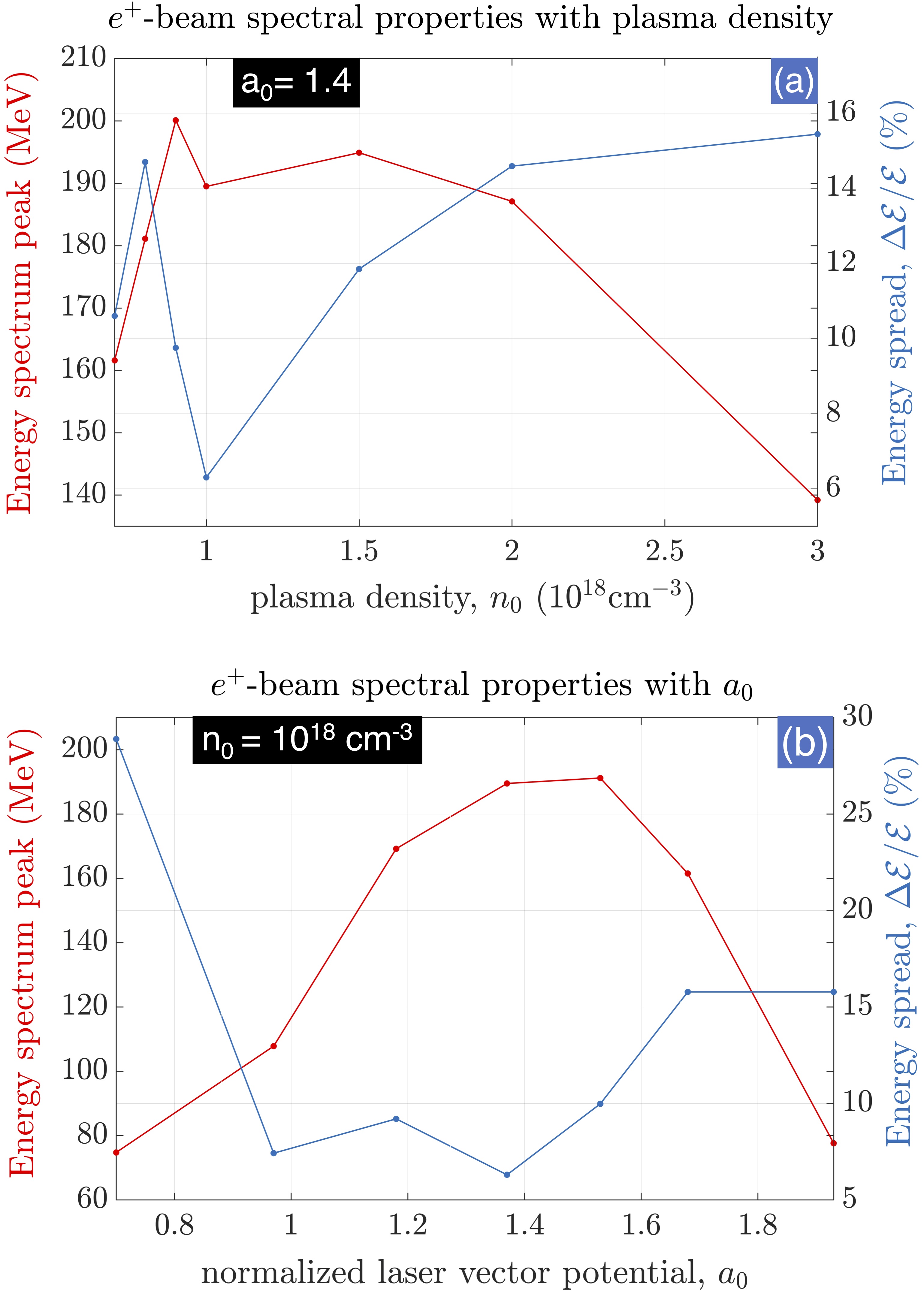}
	\end{center}
\caption{Energy spectral characteristics of Scheme-A $e^+$-beam from PIC simulations varied with $n_0$ in (a) and $a_0$ in (b).} 
\label{fig:posi-energy-variation}
\end{figure}

Multi-dimensional PIC simulations are used to validate the above analysis. Whereas 3D simulations (see Supplementary Material) offer precision,  parameter scans based on them demand inaccessible computational resources. Nevertheless, 2$\frac{1}{2}$D simulations adjusted to match 3D simulations allow for characterization over a wide parameter space. Here a 2D cartesian grid which resolves $\rm\lambda_0=0.8\mu m$ with 25 cells in the longitudinal and 15 cells in the transverse direction tracks a linearly-polarized laser pulse at its group velocity. The above detailed particle shower model is initialized as shown in Fig.\ref{fig:posi-energy-spectrum}(a),(c) (\ref{fig:posi-energy-spectrum}(e),(g)) for scheme-A (scheme-B). The shower transversely has $\sigma_r = 25\mu m$ and longitudinally spans the entire box. Each particle species is initialized with 4 particles per cell. Absorbing boundary conditions are used for both fields and particles. The laser with a Gaussian envelope of length 50fs and $a_0{\rm(2D)} =2 a_0$ propagates in $\rm50\mu m$ of free-space before it enters a fixed-ion plasma.

\begin{figure}[!hb]
	\begin{center}
   	\includegraphics[width=0.6\columnwidth]{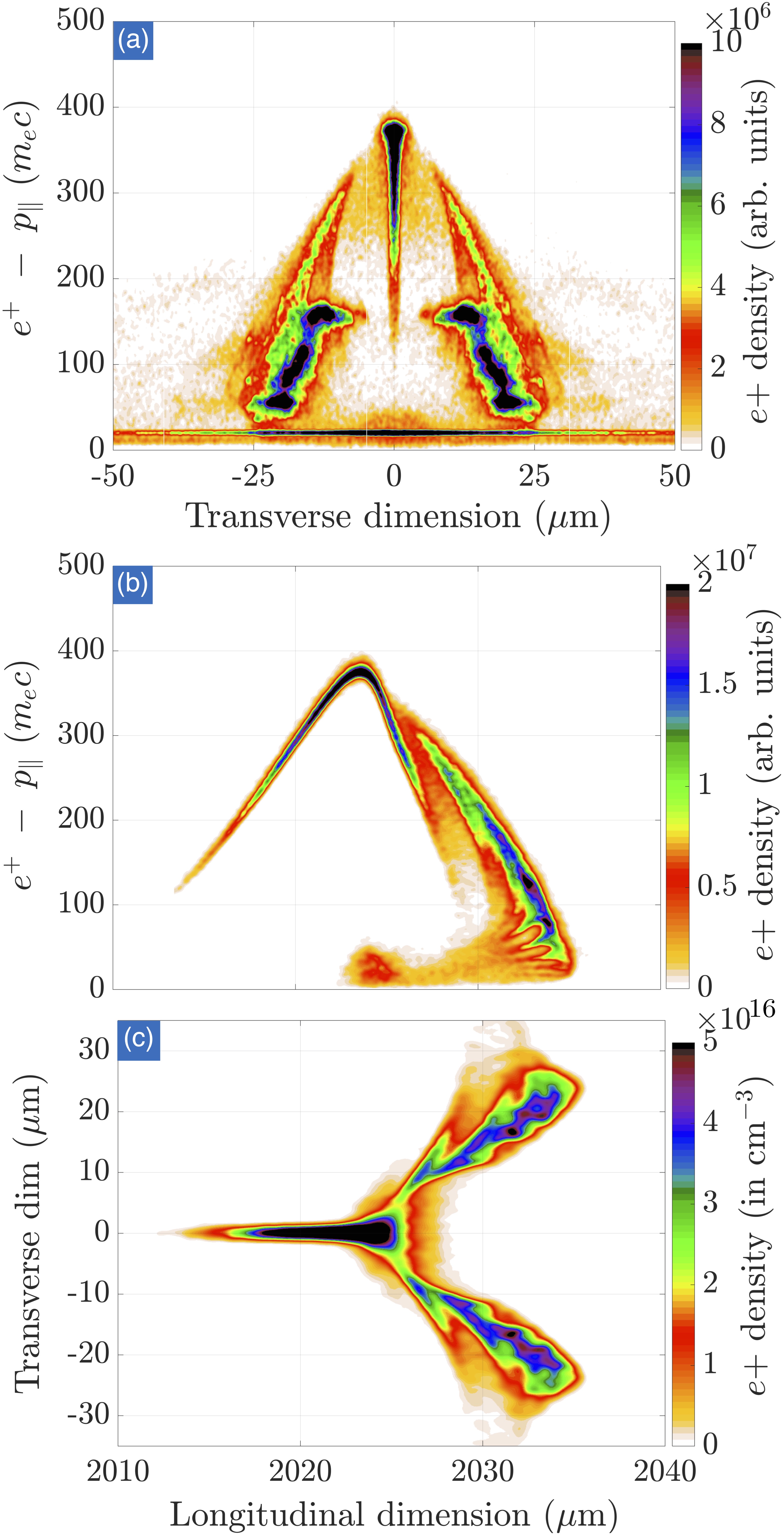}
	\end{center}
\caption{Phase-space slices of scheme-A $e^+$-beam corresponding to Fig.\ref{fig:posi-energy-spectrum}(c,d) - (a) $p_{\parallel}$-transverse space, (b) $p_{\parallel}$-longitudinal space, (c) transverse-longitudinal space. } 
\label{fig:posi-phase-spaces}
\end{figure}

The results in Fig.\ref{fig:posi-energy-spectrum} and \ref{fig:posi-energy-variation} imply that the $e^+$-LPA modeled here accelerates quasi-monoenergetic $e^+$-beams. The wave focusing fields segregate the $e^+$-beam from electrons (see Supplementary Movies). Over an $\rm L_{acc} \simeq 2mm$, $\sim$200MeV $e^+$-beams of around 5\% energy spread ($\Delta\mathcal{E}/\mathcal{E}$) are accelerated using a quasi-nonlinear wave excited by a 6J laser with 40$\rm\mu m$ FWHM spot-size in $\rm n_0=10^{18}cm^{-3}$ as shown in Fig.\ref{fig:posi-energy-spectrum}(b) (\ref{fig:posi-energy-spectrum}(f)) for scheme-A (scheme-B). These laser parameters chosen here in consideration of staging requirement of reflection off a plasma mirror \cite{electron-staging-2016}. Whereas using $\rm\Delta W=\rm\langle E_{\parallel} \rangle_{\rm L_{acc}}{\rm L_{acc}}$ the estimated $\rm \langle E_{\parallel} \rangle_{\rm L_{acc}}\simeq100GVm^{-1}$ \cite{RMP-2009} is in excellent agreement with 100 $\rm \sqrt{n_0(10^{18}\rm cm^{-3})} GVm^{-1}$, at $\gamma_{\phi}\simeq42$, $\rm L_{acc}$ is well below the de-phasing length  \cite{RMP-2009} and thus severely limits $\rm\Delta W$. This limit on $\rm L_{acc}$ is due to changes in laser properties during acceleration which modify the potential profile and the accelerating phase velocity and thrust the beam into defocusing ion-cavity phase resulting in particle loss. This limit nonetheless motivates further work to better the energy gain, energy spread and bunch charge.

The $e^+$-beam confinement properties as inferred from the phase-space slices of scheme-A $e^+$-beam in Fig.\ref{fig:posi-energy-spectrum}(c,d) shown in Fig.\ref{fig:posi-phase-spaces} are remarkable. Bunch transverse size with $\sigma_r = 5\mu m$ and length with $\sigma_z = 7.5\mu m$ are estimated from Fig.\ref{fig:posi-phase-spaces}(a,c) and (b,c), respectively. These bunch properties are consistent with eq.\ref{eq:wave-potential-lab-frame} and \ref{eq:transverse-escape-momentum}. From eq.\ref{eq:wave-potential-lab-frame} a threshold potential of $\rm \psi_{th}=0.25$ at $\rm n_0=10^{18}cm^{-3}$ is required to trap positrons upto $\rm\mathcal{E}_{sh} \geq 0.5MeV$. This value of $\rm \psi_{th}$ exceeds the eq.\ref{eq:transverse-escape-momentum} transverse threshold with $\alpha=2.5$. The observed bunch sizes are in excellent agreement with $\rm L_{\Delta}$ and $\rm R_{\Delta}$ of the simulated wave potential profiles. From $p_{\perp}$-$p_{\parallel}$ slice in Fig.\ref{fig:posi-energy-spectrum}(c), the estimated opening angle of $\sim$15mrad is perfectible. From the real space in Fig.\ref{fig:posi-phase-spaces}(c) an acceptable charge of 0.5-5pC is calculated. 

The variation of $\rm n_0$ at a fixed $a_0=1.4$ summarized in Fig.\ref{fig:posi-energy-variation}(a) implies that there is an optimal $n_0$ for a given intensity at which the peak beam energy maximizes and the energy spread minimizes. This optimality around $\rm n_0=10^{18}cm^{-3}$ is found to be due to the maximization of $\rm L_{acc}$ for the chosen laser parameters. At densities lower than the optimal smaller fields lead to slower energy gain, $\partial {\rm \Delta W}/\partial z$ and weaker beam confinement while at higher densities the laser self-focuses too rapidly. In Fig.\ref{fig:posi-energy-variation}(b), $a_0$ is varied at fixed $\rm n_0=10^{18}cm^{-3}$. An optimal quasi-nonlinear wave is only excited around $a_0=1.4$ for the above shower and laser properties. At lower $a_0$ values the initially trapped positron number is small while for $a_0$ values higher than the optimal, steepened wave ($\rm L_{\Delta}\rightarrow 0$) accelerates beams with Maxwellian energy spectra.

In conclusion, this work elucidates that control of particle-shower plasma-wave interaction, within certain limits identified here, enables all-optical acceleration of hundreds of MeV quasi-monoenergetic $e^+$-beams with pC charge using existing lasers. Future work will experimentally validate this $e^+$-LPA model, advance and explore novel high-energy antimatter applications and conceive new schemes to overcome the key limits on $e^+$-beam properties that have been identified in this work.

\begin{acknowledgments}
This work was supported by the John Adams Institute for Accelerator Science. The use of EPOCH PIC code developed in the UK is acknowledged. The simulations were initially performed using the Chanakya server at Duke University and subsequently using Imperial College High Performance Computing systems.
\end{acknowledgments}

\renewcommand{\refname}{} 

\end{document}